# Children of the Stars

*"A star is drawing on some vast reservoir of energy by means unknown to us. This reservoir can scarcely be other than the subatomic energy which, it is known, exists abundantly in all matter; we sometimes dream that man will one day learn how to release it and use it for his service."*

Arthur Eddington, 1920

**Henri Boffin**   *European Southern Observatory*

Even if it tends to hide more often in the current autumnal season, our host star, the Sun, is the principal source of energy on our planet. It has a luminosity of $3.828 \times 10^{26}$ Watts, and despite that we receive only 2 parts per billion of this, it allows for the Earth's life to thrive. Moreover, we know that the Sun has been shining in the same way for about 4.6 billion years (the age of the Earth) and will likely still do the same for another 5 billion years or so. The Sun's power engine, as well as the one of all the stars we see in the night sky, has for long been a mystery, but astronomers now have a good understanding of it.

## Explaining the Sun's energy squandering

Astronomers knew since long that there aren't many sources that could supply such amount of energy. One of the most dominant in the Universe is gravitation – the force that matters exert, and in 1856 the German physicist Hermann von Helmholtz proposed that the energy in stars was due to them contracting under the effect of gravity. But if this could produce the luminosity of a star, it can do so for only a relatively short timescale: Lord Kelvin showed soon thereafter that the Sun could only shine the way it does for 20 million years (a timescale that is now called the Kelvin–Helmholtz or thermal time scale), almost a factor 500 too small! The solution came in 1920 when Arthur Eddington realised that one of Einstein's famous equations, $E=mc^2$, could be the "the source of the heat which the Sun and stars are continually squandering." Indeed, as it had just been shown that the mass of the helium atom was less than the sum of the masses of the four hydrogen atoms which enter into it, one could then speculate that the energy was due to this difference in mass.

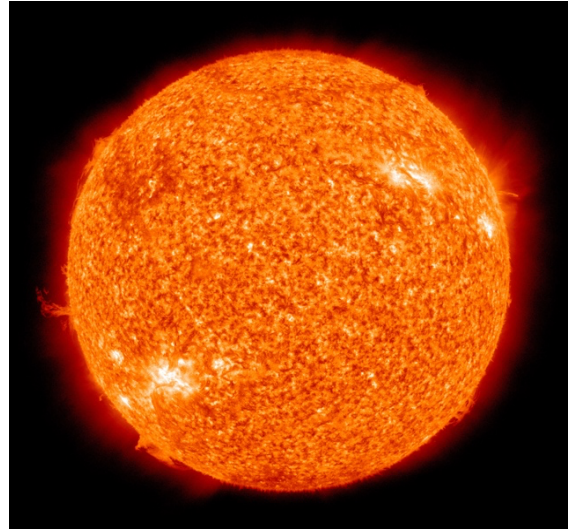

Figure 1. Our Sun is powered by fusion.
Credit: NASA/SDO (AIA)

Eddington therefore concluded that "if 5 per cent of a star's mass consists initially of hydrogen atoms, which are gradually being combined to form more complex elements, the total heat liberated will more than suffice for our demands, and we need look no further for the source of a star's energy," a most remarkable conclusion. And we now know that every second the Sun fuses in its core about 600 million tons of hydrogen into helium, thereby converting 4 million tons of matter into heat and photons, liberating as much energy as a million times what we consume on Earth every year! At this incredible rate, the Sun could in principle live for almost 80 billion years, but as only the most central parts can be used, the real lifetime of the Sun on this reaction is about 10 billion years.

## A drunken walk

It is rather easy to estimate the temperature at the centre of the Sun: we just need to equate the thermal energy of a proton to the gravitational energy and this tells us that deep inside the Sun, the temperature is 15 million K, and the more massive a star, the hotter its core is. This also tells us that the photons that escape from the Sun's surface do not come directly from its centre, as such a temperature correspond to gamma and X-rays and not the yellow light that mostly emanates from the Sun. The reason for this is that the path of a photon inside the Sun isn't straight! Instead, every tenths of a millimetre or so, and this over the 700,000 km of the Sun's radius, the photon will encounter a particle, and will either be scattered, or absorbed and reemitted at lower energy – a process that resembles the path of a drunk person and that is called a random walk. It will on average take between 20,000 and 200,000 years for a photon that was produced in the core of the Sun to emerge!

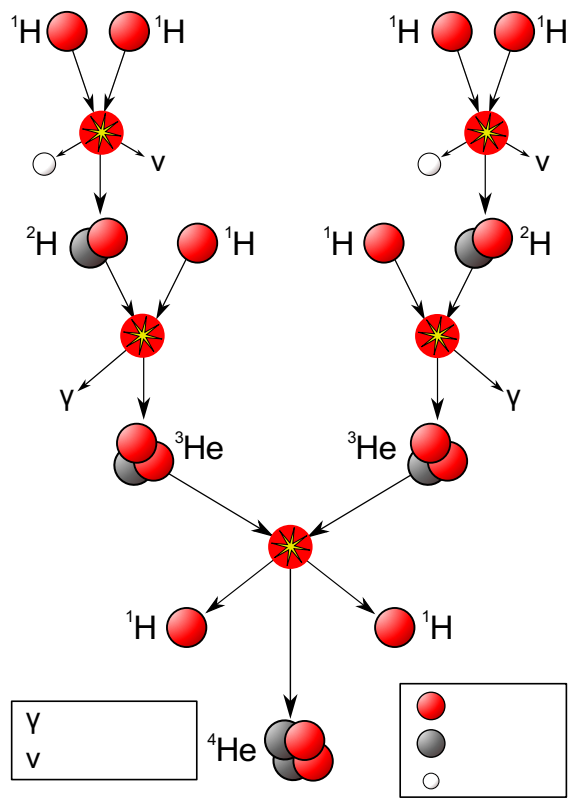

Figure 2. The p-p chain.
Credit: Borb/Wikipedia

After Eddington brilliant assertion, it took still some more years to learn the intricate details of the processes. First, it is important to realise that most of the protons inside the Sun don't move fast enough to penetrate the electromagnetic Coulomb barrier that would allow their fusion, and so only the quantum mechanics tunnelling effect can explain why this works. Thus, contrarily to what is often thought, a proton can live at the centre of the Sun for almost 3 billion years before it will fuse with another proton. This explains why stars like the Sun live so long, allowing for life (and us) to come along.

### Finally, the full explanation

It is in 1938, finally, that after some pioneering work by George Gamow, Robert d'Escourt Atkinson, and Carl Friedrich von Weizsäcker, among others, that Hans Bethe discovered the exact processes that can combine hydrogen into helium at the centre of stars. He would receive the Nobel Prize in Physics for this in 1967. Bethe showed that there were two possible processes to fuse four nuclei of hydrogen into one of helium. The first one is called the p-p chain (for proton-proton; Fig. 2), even if the fusion of two protons is only the beginning of a much more complicated process: the two protons form deuterium (or "heavy hydrogen"), which fusing with another proton, a process that takes only 1.4 seconds on average, will lead to helium-3. Finally, two of these helium-3 nuclei will merge, every 240 000 years or so, to produce the stable helium-4 nucleus and two protons, which can be reused, at the same time as releasing energy and a vast number of neutrinos. Every second, $2\times10^{38}$ neutrinos are produced in the Sun – a fraction of which will pass right through the Earth (and about $10^{12}$ through each of us!) without being at all affected. Some are recorded by dedicated experiments buried deep in the ground though and these have shown that this is indeed the process that must be at play at the centre of our host star. The detections of these solar neutrinos led to another Nobel prize in 2002. As the p-p chain requires temperatures above about 3 million K in the centre of a star to start, this also limit the mass of what can become a star: this can be computed as 0.08 times the mass of the Sun. Objects which are less massive than that cannot start hydrogen burning in their core and are thus 'failed stars'. They are called *brown dwarfs*.

The second way to produce helium from protons is through the six-stage carbon-nitrogen-oxygen (CNO) cycle (Fig. 3). If stars are made mostly from hydrogen and helium, they also contain some small fraction of heavier elements, including carbon, nitrogen or oxygen. As Hans Bethe showed, these can add as catalyst in the formation of helium. This process, however, requires higher temperature and is therefore dominating the energy production rate of stars more massive than about 1.3 times the mass of the Sun. In the Sun, it is thought that around 1% of all its energy is produced by the CNO cycle. Only very recently, the international Borexino collaboration directly detected neutrinos produced in the CNO cycle, providing the first direct proof of this process.

### The future of the Sun

What will happen in about 5 billion years, when the hydrogen in the core of the Sun will have been exhausted? As gravity rules, the core of the Sun will gradually contract and become hotter, while the outer layers of the Sun will expand – the Sun will become a red giant star, becoming 2,300 times more luminous and about 170 times larger than it is now, thereby engulfing Mercury and Venus and likely melting the Earth's surface. When the core's will have contracted enough that the central temperature reaches about 100 million K, helium will be ignited and transformed into carbon and oxygen.

The way this is partly done, via the 'triple-alpha process' (Fig. 4) is another testimony of the genius of some astronomers. This process is a two-step reaction in which a carbon nucleus is formed from three helium nuclei. First, two alpha-particles (helium nuclei) collide to form a beryllium-8 nucleus. The latter is unstable and decays so rapidly that it is unlikely that a third helium nucleus could be captured before the beryllium-8 decays. To create significant amounts of carbon in the Universe, an additional factor that makes a successful combination more likely would be needed. And we know that carbon is created – if it were not, then we humans, and all life on Earth, would not be here to discuss it, and you would not be reading this article! Using this simple but profound argument, the remarkable British astronomer Fred Hoyle, in a characteristic stroke of genius, predicted that some additional helping factor must indeed exist. Scientists turned to laboratory experiments and, sure enough, discovered a previously unknown 'resonance', a matching of energy levels between the beryllium-8 and helium-4 nuclei, and the carbon-12 nucleus that they form. This resonance greatly increases the probability of a successful combination, just as Hoyle predicted. Thus, helium will be transformed into carbon and some of the latter will also fuse with another helium nucleus to form oxygen. This will be the last stage for the Sun which will evolve to become a white dwarf. Stars more massive than eight times the Sun, on the other hand, will be able to continue the adventure, producing heavier elements until the most stable of all, iron.

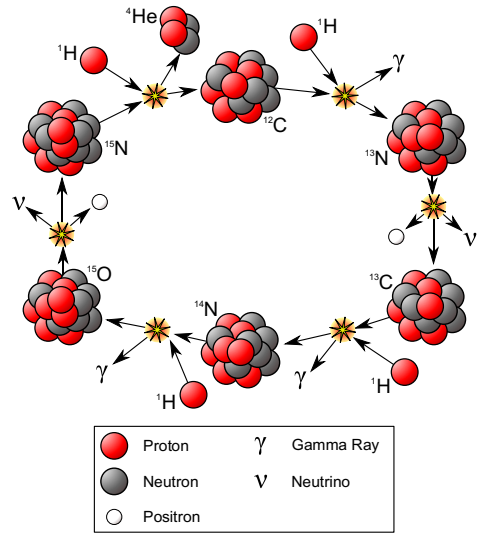

Figure 3. The CNO cycle.
Credit: Borb/Wikipedia

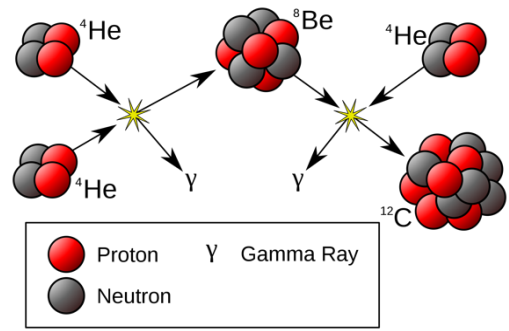

Figure 4. The triple-alpha process to convert helium into carbon. Credit: Borb/Wikipedia

We now know that apart from the lighter elements – hydrogen and helium, and some of the lithium – which were created during the 'Big Bang', all the elements from carbon to uranium have been produced by nuclear processes in the interior of stars or their explosions: humans, all other living creatures on Earth – and Earth itself – are children of the stars! So, it is also quite remarkable that the CNO that is used by the more massive stars to fuse hydrogen into helium was created in previous generations of stars as well. The Universe is a most remarkable recycling machine!